# Broadband photonic RF channelizer with a Kerr soliton crystal micro-comb

Mengxi Tan[1], Xingyuan Xu[1,2], and David J. Moss[1], *Fellow IEEE*

[1]Optical Sciences Centre, Swinburne University of Technology, Hawthorn, VIC 3122, Australia
[2]Present Address: Dept. of Electrical and Computer Systems Engineering, Monash University, Clayton, 3800 VIC, Australia

*Abstract*— **We report a 92 channel RF channelizer based on a 48.9 GHz integrated micro-comb that operates via soliton crystals, together with a passive high-Q ring resonator that acts as a periodic filter with an optical 3dB bandwidth of 121.4 MHz. We obtain an instant RF bandwidth of 8.08 GHz and 17.55 GHz achieved through temperature tuning. These results represent a major advance to achieving fully integrated photonic RF spectrum channelizers with reduced low complexity, size, and high performance for digital-compatible signal detection and broadband analog signal processing.**

## I. INTRODUCTION

Modern RF systems require receivers as key components for electronic warfare, radar systems, satellite communications, and more [1-4], since they need to analyze and detect RF signals. However, the immediately accessible bandwidth (ie., without adjusting or thermal tuning for example) for systems to receive RF signals exceeds the capability of typical analog-to-digital converters (ADCs). Therefore, detecting and analyzing signals using flexible and powerful digital-domain tools requires that the broadband signal be sliced spectrally into segments that are compatible with digital systems, [5] and this is the purpose of RF channelizers. While electronic based RF channelizers are limited by the electronic bottleneck in bandwidth, approaches based on photonics are attractive since they can provide very large bandwidths as well as low loss, together with very strong electromagnetic interference immunity.

A substantial body of work has been devoted towards developing RF channelizers based on photonics. These can generally be categorized into two different approaches. The first approach operates through separating the RF spectrum into channels with many high resolution and accurately spectrally positioned narrow passband optical filters [6]. While this is effective, it requires a larger footprint and cannot achieve extremely high RF resolution or bandwidth. The other approach to RF channelizers based on photonic uses a much more efficient and compact method that relies on a Vernier shift between a source of multiple wavelengths and an optical filter that is spectrally periodic, in order to realize an accurately progressively stepped broadband RF spectral channelization. A range of different approaches have been based on this strategy including stimulated Brillouin scattering [7], nonlinear optics in fibers [8], incoherent spectrally sliced sources [9], discrete arrays of laser diodes [10], electro-optic modulator generated frequency combs [5, 11], and others [12]. However, many of these suffer from drawbacks of one form or another including a limited number of channels, limited optical resolution, and difficulty in integrating the system due to the need discrete components.

Microcombs [13-17] have recently attracted significant attention, including CMOS compatible devices [18 - 23], since they can provide a large number of coherent channels in a very small (square millimeter) chip. They have formed the basis of a wide array of RF demonstrations [34-48, 35] including true RF time delays [36, 37], transversal filter based signal processing [38 - 43], RF frequency mixing and conversion [44], signal generators of various forms including phase-encoded generators [45], RF channelizers [47, 48] and many others [49-58]. We recently reported [47] a microcomb based RF channelizer that exhibited very high performance, where a 200 GHz FSR microcomb provided twenty wavelengths across the C band. This microcomb was then filtered with a passive ring resonator with a FSR spacing of 48.9GHz. The RF signal was multicast onto the 200GHz microcomb which was subsequently sliced by every fourth passive ring resonator resonance. The net difference between the FSR of the microcomb and 4x the passive ring filter produced a frequency offset of 4.4 GHz. This resulted in the RF frequency step between the different RF channels, and since it was much larger than the 120MHz channel resolution, did not allow instantaneous spectral coverage of the full continuous RF spectrum. This resulted in a discontinuous RF operation band of 4 channels that could operate at one time, and so the overall bandwidth, given by the product of the slicing resolution and the channel number, was limited with "gaps". These gaps, however, could be filled in via temperature tuning the passive ring. The comparatively large 200GHz comb FSR yielded only 20 wavelengths across the C-band.

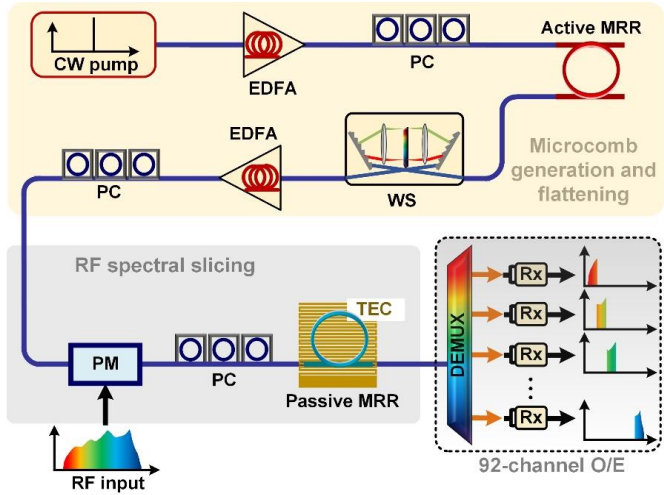

Fig. 1. Schematic of the broadband RF channelizer based on a soliton crystal microcomb. EDFA: erbium-doped fibre amplifier. PC: polarization controller. MRR: micro-ring resonator. WS: Waveshaper. PM: phase modulator. TEC: temperature controller. DEMUX: de-multiplexer. Rx: Receiver.

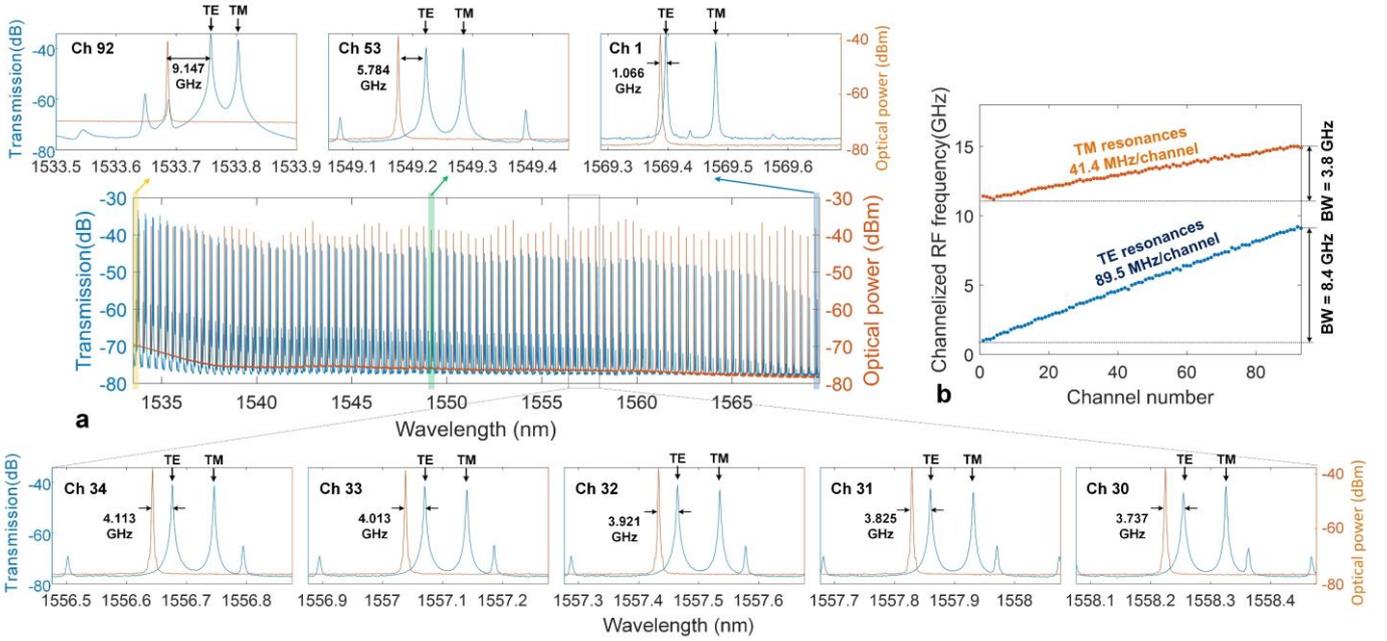

Fig. 2. (a) The measured optical spectrum of the micro-comb and drop-port transmission of passive MRR. (b) Extracted channelized RF frequencies of the 92 channels, calculated from the spacing between the comb lines and the passive resonances. Note that the labelled channelized RF frequencies in (a) is adopted from the accurate RF domain measurement using the Vector Network Analyzer.

In this paper, [48] we demonstrate an RF channelizer based on photonics that achieves higher performance, accomplished through the use of two ring resonators with small, and closely matched, FSRs, at 49GHz. The first MRR generated a microcomb that operated via soliton crystals, while the second resonator acted as a passive filter, providing two major benefits. First, the very small FSR of 49GHz yielded 92 wavelengths across the C band which significantly increased the instant bandwidth to 8.08 GHz – which is over 22x more than the previous device [47]. By using a ring with a closely matched FSRs to filter RF spectrum using Vernier shifting, a frequency step of the RF channelization \of 87.5 MHz was demonstrated. This led to a continuous channelization spectrum of the RF signal because the channelization step was less than the 121.4 MHz optical resolution. The other major advantage of our approach is that we exploited parallel conversion of phase to intensity-modulation (PM - IM) for all wavelengths simultaneously — which provided an instant RF output, while still in a stable and very small footprint without needing local oscillators. Lastly, the frequency range of operation of the channelizer could be tuned dynamically by varying the microcomb to passive ring offset frequency. We accomplished this by temperature tuning the passive ring, thus demonstrating RF channelization across a 17.55 GHz total frequency range. This approach achieved high performance in a small footprint with low potential cost and complexity.

## II. OPERATION PRINCIPLE

Figure 1 illustrates the operational principle of the RF channelizer. It comprises three sections - first the is produced and spectrally flattened with an optical waveshaper. With $N$ comb lines at a $\delta_{OFC}$ spacing, the $k^{th}$ ($k$=1, 2, 3, …, 92) optical comb frequency is

$$f_{OFC}(k) = f_{OFC}(1) + (k-1)\delta_{OFC} \quad (1)$$

with $f_{OFC}(1)$ being the first comb line frequency on the long wavelength (red) side.

In the second module, the flattened comb lines were passed through an electro-optic phase modulator in order to multicast the RF signal onto all wavelengths. In the last module Finally, the replicated RF spectra are sliced by the passive MRR with an FSR of $\delta_{MRR}$, with the resolution given by the 3dB bandwidth of the resonator. As a result, the different RF channels have a center frequency given by

$$f_{RF}(k) = f_{MRR}(k) - f_{OFC}(k)$$
$$= [f_{MRR}(1) - f_{OFC}(1)] + (k-1)(\delta_{MRR} - \delta_{OFC}) \quad (2)$$

where $f_{RF}(k)$ is the $k_{th}$ channelized RF frequency, and $f_{MRR}(k)$ is the passive ring's $k^{th}$ center frequency. Here, the relative difference between the 1st comb line and adjacent passive ring resonance is given by $[f_{MRR}(1) - f_{OFC}(1)]$, which yields the offset RF channel frequency, and $(\delta_{OFC} - \delta_{MRR})$ is the RF frequency step between adjacent wavelengths.

We used a combination of notch filtering with phase modulation to perforrm phase to intensity modulation conversion. Phase modulating first generated both lower and upper sidebands that had opposite phase. One of the sidebands was then eliminated by the notch filter resonance, which left the other unsuppressed optical sideband to beat with the optical carrier after detection. Thus, this approach directly converted the modulation format from phase to single sideband intensity, providing $N$ parallel bandpass filters with a spectral resolution $\Delta f$ determined by the linewidth of the passive ring. The RF center frequencies $f_{RF}(k)$ were given by the offset between the passive ring resonant frequencies and the comb lines $[f_{MRR}(k) - f_{OFC}(k)]$, given by Eq. (2). As a result, the input RF spectrum was channelized into $N$ segments, each centered at $f_{RF}(k)$ and with a bandwidth of $\Delta f$. This approach did not require local oscillators to perform coherent homodyne detection, and so it was much more stable and compact [48].

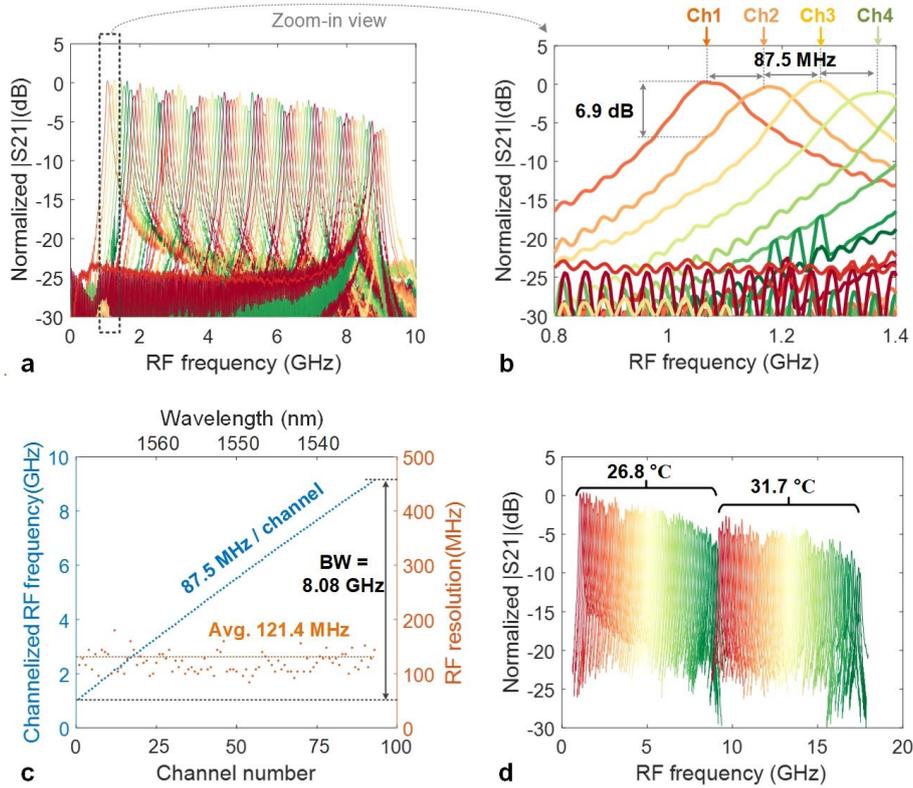

Fig. 3. RF transmission spectra of (a) the 92 channels and (b) a zoom-in of the first 4 channels. (b) Extracted channelized RF frequency and resolution. (d) Measured RF transmission spectra vs chip temperature of the passive MRR.

Finally, the wavelengths were separated and converted into the electric signals by separate photodetectors, thus yielding $N$ simultaneous RF channels with a spectral width of $\Delta f$, which is in the range of ADCs. After that, the channelized RF signals are converted into digital signals via an array of ADCs and processed with digital domain tools for further analysis.

## III. RESULTS

The ring resonators were fabricated in Hydex which is CMOS compatible, and featured low propagation loss of 0.06 dB per cm, a moderately high nonlinearity of 233 /W / km) and no measurable nonlinear loss [13, 14]. The rings had the same radius of 592 μm with a 1.5×2 μm cross sections and Q's of greater than 1 million, with a total fibre to fibre insertion loss of 1.5 dB. The active ring featured anomalous dispersion in the C band to support parametric gain, as well as having a mode crossing near 1556 nm, needed to generate soliton crystal microcomb. The power of the pump was amplified to 2W and then its wavelength was manually swept from blue to red to initiate the comb [59, 60], that had a palm-like spectrum with a 48.9 GHz spacing, producing 92 wavelengths across the C-band. The microcomb was low-noise, coherent, and very easy to generate. All 92 comb lines were flattened and passed through the phase modulator, driven by the RF signal in order to multicast the RF signal onto the optical carriers at all wavelengths. The RF replicas were then sliced spectrally via the passive ring.

The passive ring transmission spectrum is shown in Fig. 2(a) measured with a broadband EDFA source. The RF channelized frequency $f_{RF}(k)$, determined by the comb line to adjacent passive resonance spacing, decreased monotonically from blue to red. The total RF channelized frequency (Fig. 2(b)) had a 8.4 GHz bandwidth for TE and 3.8 GHz for TM, with steps of 89.5 (TE) and 41.4 (TM) MHz per channel. The channel resolution was 120 MHz for TE polarization. The polarization was aligned to the TE mode of the passive ring and so the TM modes were not coupled into. The RF spectrum was then channelized into many segments over different wavelengths, which were separated into parallel spatial paths and detected individually. We used a Vector Network Analyzer to measure the RF transmission spectrum (Fig. 2 (a, b)). The channelizer SNR was 23.7 dB, which can be improved by increasing the rejection of the passive optical ring filter. This can be achieved by optimizing the coupling coefficient for the bus waveguide to ring resonator.

The RF channel center frequencies (Fig. 3) had a 87.5 MHz per channel step with an overall 8.08 GHz instant bandwidth, agreeing with calculations based on the ring's spectrum. The average RF channel resolution obtained from the ring 3dB bandwidth for the RF channel's transmission spectrum was 121.4 MHz. This high a resolution proves that this method is compatible with the modest bandwidths of most digital components. Finally, the operation band of the channelizer was variable from 1.006-9.147 GHz to 9.227-17.49 GHz, by temperature tuning the passive ring on millisecond timescales thus reaching an overall RF bandwidth of 16.48 GHz (Fig. 3(d)). The maximum channelizer operation frequency (Nyquist zone) was 48.9 GHz/2 = 24.45 GHz. Larger spaced microcombs can increase the maximum RF to 100 GHz, but this comes with a tradeoff in the comb line number.

## IV. CONCLUSION

We report a RF channelizer with high performance based on Kerr microcombs that operate via soliton crystals. The 49GHz FSR of both the passive and active rings that closely matched each other produced 92 wavelengths over the C-band. This enabled a very large 8.08 GHz instant RF bandwidth and maximum range of 17.49 GHz, achieved via temperature tuning. The channelizer had a high RF resolution of 121.4 MHz. This method is highly attractive to achieve instantaneous broadband RF signal detection and processing, towards photonic integrated RF system receivers.